\documentstyle[psfig,epsf]{mn}
\newcommand{\ltaraw}{$\; \buildrel < \over \sim \;$}
\newcommand{\lta}{\lower.5ex\hbox{\ltaraw}}
\newcommand{\gtaraw}{$\; \buildrel > \over \sim \;$}
\newcommand{\gta}{\lower.5ex\hbox{\gtaraw}}

\newcommand{\ffffff}[1]{\mbox{$#1$}}
\newcommand{\scnd}{\mbox{\ffffff{''}\hskip-0.3em.}}
\newcommand{\scmd}{\mbox{\ffffff{''}}}

\loadboldmathitalic 
\title [SN1997ff and Gravitational Lensing]
{Was SN1997ff at $z\sim1.7$ magnified by gravitational lensing?}
\author[G. F. Lewis \& R. A. Ibata]
{Geraint F. Lewis$^{1}$ \& R. A. Ibata$^{2}$\\
$^{1}$
Anglo-Australian Observatory, P.O. Box 296, Epping, NSW 1710, Australia:
Email \tt{gfl@aaoepp.aao.gov.au}\\
$^{2}$
Observatoire de Strasbourg, 11, rue de l'Universit\'e, F-67000, Strasbourg, 
France:
Email \tt{ibata@astro.u-strasbg.fr}
}
\date{\today}
\begin{document} 
\maketitle 
\begin{abstract}
The quest  for the cosmological  parameters has come to  fruition with
the  identification  of  a  number  of supernovae  at  a  redshift  of
$z\sim1$. Analyses of the  brightness of these standard candles reveal
that the Universe is dominated  by a large cosmological constant.  The
recent  identification  of the  $z\sim1.7$  SN1997ff  in the  northern
Hubble Deep  Field has provided  further evidence for  this cosmology.
Here we examine the case  for gravitational lensing of SN1997ff due to
the presence of galaxies lying along  our line of sight. We find that,
while  the  alignment  of  SN1997ff  with foreground  masses  was  not
favorable for  it to  be multiply imaged  and strongly  magnified, two
galaxies did lie close  enough to result in significant magnification:
$\mu\sim1.4$  for  the  case  where  these  elliptical  galaxies  have
velocity  dispersion $200  {\rm  km/s}$.  Given  the small  difference
between  supernova  brightnesses  in different  cosmologies,  detailed
modeling of  the gravitational  lensing properties of  the intervening
matter is therefore required before the true cosmological significance
of SN1997ff can be deduced.
\end{abstract}

\newcommand{\sn}{SN1997ff}

\section{Introduction}\label{introduction}
The search for the values  of the cosmological parameters has occupied
astronomers  for almost  a century.   Recently, a  study of  the light
curves of local supernovae Ia  has provided an accurate calibration of
their intrinsic  luminosity (Hamuy et  al.  1996; Riess et  al. 1998).
With  such a  standard candle  in  hand, the  measurement of  apparent
supernovae  brightnesses  at   high  redshift  allows  the  underlying
cosmological parameters to  be determined.  This has been  the goal of
two teams who have now discovered approaching one hundred supernovae out to
a redshift of $z\sim1.2$ (Schmidt et al. 1998; Perlmutter et al. 1999;
Riess et al. 2000).  The results of these studies suggest that we live
in  a universe  which is  dominated by  a cosmological  constant, with
$\Lambda_o=0.7$,  with enough matter,  $\Omega_o=0.3$, to  flatten the
overall  topology,  as  seen   in  studies  of  the  cosmic  microwave
background (Melchiorri et al. 2000).

With the release  of the cosmological supernovae results,  a number of
analyses of physical processes that could also dim distant supernovae,
and therefore  mimic a cosmological constant,  were considered.  These
included   gravitational   lensing   (e.g.    C{\'e}l{\'e}rier   2000;
Bergstr{\"o}m, Goliath, Goobar, \& M{\"o}rtsell 2000), dust (Totani \&
Kobayashi  1999; Croft,  Dav{\'e}, Hernquist,  \& Katz  2000)  and the
evolution  of  supernova properties  (H{\"o}flich,  Nomoto, Umeda,  \&
Wheeler 2000);  see Riess  (2000) for a  more complete  description of
these effects.  Recently, however, the  discovery of a supernova  at a
redshift    of    $z\sim1.7$   has    been    announced;   see    {\tt
http://oposite.stsci.edu/pubinfo/pr/2001/09}.  Located in the northern
Hubble  Deep Field  (Williams et  al.  1996;  Ferguson,  Dickinson, \&
Williams 2000), this supernova, designated \sn, was identified via the
comparison of  exposures at differing epochs by  Gilliland, Nugent, \&
Phillips (1999).  While a  publication detailing the supernova has yet
to  appear,  the press-release  announces  that  this  system, with  a
photometric  redshift of $z=1.7\pm0.15$,  represents the  most distant
supernova discovered and that it confirms the earlier conclusions of a
cosmological constant dominated universe, as it is brighter than would
be expected in  a simple matter dominated universe.   Such dimming and
brightening,   relative  to   matter  dominated   cosmologies,   is  a
peculiarity  expected  in universes  with  a substantial  cosmological
constant, but is not consistent with the action of dust, gravitational
lensing or supernova evolution.

An examination  of the  Hubble Deep Field,  however, reveals  that the
host of \sn\  is seen in projection near  two large elliptical systems
at  a  redshift of  $z\sim0.56$.   In  this  Letter, we  consider  the
gravitational  lensing influence  of  these galaxies  on the  apparent
brightness of \sn.

\section{The Environment of \sn}\label{environ}
\begin{figure}
\centerline{ \psfig{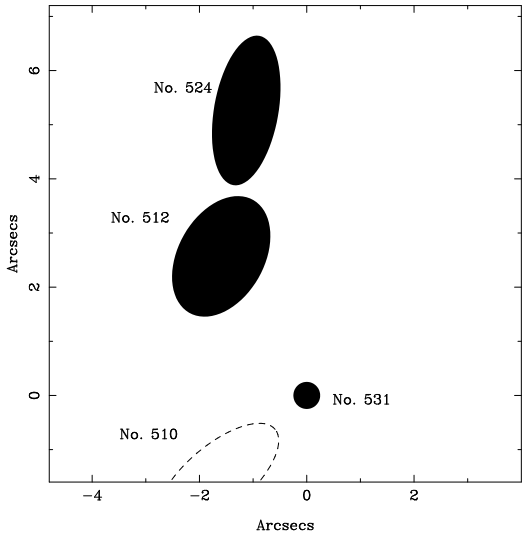} }
\caption{Schematic depiction of the  environment of the host galaxy of
\sn\ (No.~531).   Both galaxy  No.~512 and No.~524  have spectroscopic
redshifts of $z\sim0.56$, while  No.~510 has a photometric redshift of
$z\sim1.9$.}
\label{Figure1}
\end{figure}

Examining the  host galaxy of  \sn\ in the  HDF North reveals  that it
lies in  close proximity to  several other systems; this  is presented
schematically  in  Figure~\ref{Figure1}. The  labels  for the  objects
depicted are taken  from S.  Gwyn's photometric catalog  of objects in
the   Hubble  Deep   Field  (Gwyn   \&   Hartwick  1996)~\footnote{\tt
http://astrowww.phys.uvic.ca/grads/gwyn/pz/hdfn/spindex.html}.      The
host of  \sn\ is No.~531,  with a tentative spectroscopic  redshift of
$z\sim1.66$, while No.~512 and No.~524 possess spectroscopic redshifts
of $z=0.555$  and $z=0.557$ respectively  (Cohen et al.  1996). System
No.~510  possess a  photometric redshift  of $z\sim1.9$  and therefore
probably lies beyond the host of \sn.  Two other systems at a redshift
of $z\sim0.85$ lie  off the bottom right-hand corner  of the plot, but
as these are at a greater projected distance than galaxy No.~512, they
are neglected in this analysis.

\section{Gravitational Lensing}\label{gravlens}
Given their proximity to the  line-of-sight, it is prudent to consider
the gravitational lensing influence of galaxies No.~512 and No.~524 on
the  observed  brightness  of  \sn.   For  galactic  mass  objects  at
cosmological  distances,  the  typical  scale for  strong  lensing  is
$\sim1\scmd$ and, given the separation of the galaxies and the host of
\sn, a  distance of $3\scnd0$  for No.~512 and $5\scnd4$  for No.~524,
there  is  little  possibility  that  the supernova  would  have  been
multiply imaged or substantially magnified. More subtle magnifications
however, will occur outside the region of multiple imaging.

To examine this further we adopt the simple pseudo-isothermal model of
the two-dimensional  gravitational potential to represent  each of the
foreground galaxies (Kochanek,  Blandford, Lawrence, \& Narayan 1989).
These  are assumed to  be spherical  and are  centered on  the optical
positions of No.~512 and  No.~524. The normalization of this potential
depends upon  the velocity dispersion  of the lensing objects  and the
ratio  of  the  lens-source   and  observer  source  angular  diameter
distances.    Calculating  the   distances  in   various  cosmologies,
including  those   with  a  substantial  $\Lambda$   term,  using  the
algorithms of Kayser, Helbig, \&  Schramm (1997), it is found that the
normalization   is   relatively   insensitive  to   the   cosmological
parameters, with  a $<5\%$  variation between the  total magnification
for reasonable cosmologies.  It  is also assumed that galaxies No.~512
and No.~524 have the same velocity dispersion.

Three  fiducial  values of  the  velocity  dispersion are  considered,
100km/s, 200km/s and 300km/s, the central value representing a typical
elliptical galaxy  (Whitmore, McElroy, \&  Tonry 1985). For  the lower
velocity  dispersion of 100km/s,  the resulting  gravitational lensing
magnification  of \sn\  due to  galaxies No.~512  and No.~524  is only
$\mu\sim1.08$. Considering instead the  more typical value of 200km/s,
the    gravitational   lensing    magnification    is   $\mu\sim1.42$,
corresponding to  a brightening of  0.38mags. If galaxies  No.~512 and
No.~524 are massive ellipticals, with velocity dispersions of 300km/s,
the magnification is $\mu\sim2.92$, a brightening of 1.16mags.

\section{Discussion and Conclusions}\label{conclusions}
The simple  analysis presented in  this paper demonstrates that  it is
likely that the recently  identified high redshift supernova, \sn, was
mildly  magnified  by  the  gravitational  lensing  influence  of  two
galaxies  lying close  to the  line  of sight.   The magnification  is
dependent  upon the mass  of the  lensing galaxies,  but if  these are
typical elliptical galaxies, \sn\ appeared brighter by 0.38mags. While
this value will depend upon  the assumed form of the mass distribution
in  the lensing galaxy,  the estimate  of the  magnification presented
here will be of the right order.

\begin{figure}
\centerline{ \psfig{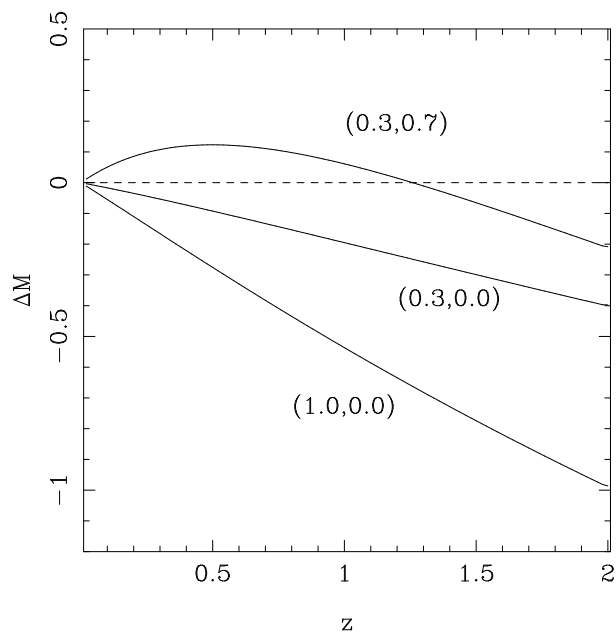} }
\caption{The difference  in apparent  magnitude of a  standard candles
between  an empty universe  (dashed line  at $\Delta$M$=0$)  and other
cosmologies.  The brackets  list the parameters ($\Omega_0,\Lambda_0$)
of the model.}
\label{Figure2}
\end{figure}

The  difference  in  apparent  brightness  of  a  standard  candle  in
different  cosmologies  is   presented  in  Figure~\ref{Figure2}.   At
$z=1.7$, the difference between  alternate world-models is slight, for
instance, the  dark energy dominated model predicts  that objects will
be  only  0.13mags brighter  than  in  an  empty universe  model.  Our
analysis suggests that gravitational lensing enhanced the magnitude of
\sn\ by $\sim  0.4$mags, an effect larger than  the difference between
those  two cosmologies.  Indeed,  in the  light  of our  gravitational
lensing analysis, the  new supernova datum may be  in closer agreement
to an  empty universe model (though  of course, our  analysis does not
affect previous  conclusions drawn from  the sample of  lower redshift
supernovae).   For the true  cosmological significance  of \sn\  to be
determined, therefore,  a detailed model of  the gravitational lensing
in this system is required.

While  \sn\ is  the first  supernova to  be identified  in  the higher
redshift  universe, it  is likely  that others  too will  be  found in
similar circumstances.   Hence, the determination  of the cosmological
parameters via supernova explosions is  likely to be come tangled with
the  problem  of  understanding   the  mass  distribution  of  lensing
galaxies, something  that has dogged  the determination of  the Hubble
Constant from gravitational lensing time  delays for a number of years
(e.g. Schechter et al. 1997).

\newcommand{\mnras}{MNRAS}
\newcommand{\nat}{Nature}
\newcommand{\araa}{ARAA}
\newcommand{\aj}{AJ}
\newcommand{\apj}{ApJ}
\newcommand{\apjl}{ApJ}
\newcommand{\apjs}{ApJSupp}
\newcommand{\aap}{A\&A}
\newcommand{\aaps}{A\&ASupp}
\newcommand{\pasp}{PASP}

\end{document}